\documentclass[3p,times,twocolumn]{elsarticle}

\usepackage{ecrc}

\volume{00}

\firstpage{1}

\journalname{Nuclear Physics B Proceedings Supplement}

\runauth{M. A. Nefedov et al.}

\jid{nuphbp}

\jnltitlelogo{Nuclear Physics B Proceedings Supplement}

\usepackage{amssymb}

\usepackage[figuresright]{rotating}

\newcommand{\re}{\mathop{\mbox{Re}}\nolimits}
\newcommand{\im}{\mathop{\mbox{Im}}\nolimits}
\newcommand{\tr}{\mathop{\mbox{tr}}\nolimits}
\newcommand{\sgn}{\mathop{\mbox{sgn}}\nolimits}
\newcommand{\Hc}{\mathop{\mbox{h.c.}}\nolimits}

\begin{document}

\begin{frontmatter}



\dochead{}

\title{Towards NLO calculations in the
parton Reggeization approach}


\author[SU]{M. A. Nefedov\fnref{speaker}}
\author[SU]{V. A. Saleev}

\tnotetext[speaker]{Speaker. E-mail addresses: {\tt nefedovma@gmail.com}, {\tt saleev@samsu.ru}}

\address[SU]{Samara National Research University, Moscow Highway, 34, 443086, Samara, Russia}
\begin{abstract}
  Parton Reggeization approach is the scheme of $k_T$-factorization for multiscale hard processes, which is based on the Lipatov's
  gauge invariant effective field theory (EFT) for high energy processes in QCD. The new type of rapidity divergences, associated
  with the $\log 1/x$-corrections, appears in the loop corrections in this formalism. The covariant procedure of regularization
  of rapidity divergences, preserving the gauge invariance of effective action is described.  As an example application,
  the one-loop correction to the propagator of Reggeized quark and $\gamma Q q$-scattering vertex are computed.
  Obtained results are used to construct the Regge limit of one-loop $\gamma\gamma\to q\bar q$ amplitude.
  The cancelation of rapidity divergences and consistency of the EFT prediction with the full QCD result is demonstrated.
  The rapidity renormalization group within the EFT is discussed.
\end{abstract}


\end{frontmatter}


\section*{Introduction.}




 The modern theoretical description of hard processes in hadronic collisions is based on the collinear factorization
  of the cross-section into the parton distribution functions (PDFs) $f_i(x,\mu^2)$ and hard-scattering coefficient
   $d\hat{\sigma}(x_1,x_2,Q,\mu)$. PDFs depend on the typical momentum scale of the hard subprocess ($Q\gg \Lambda_{QCD}$)
    and the fraction ($x$) of the momentum of the proton, carried by the parton.  Momenta of partons in the initial state
     of the hard-scattering subprocess are parametrized as $q_{1,2}^\mu=x_{1,2}P_{1,2}^\mu$, so that $q_{1,2}^2=0$~\cite{Collins}.
      Such description is well suited for the processes with the single hard scale $Q$.

 For the case of the processes with two widely separated scales $Q_1\gg Q_2\gg \Lambda_{QCD}$, large logarithmic corrections,
 proportional to the $\log Q_1/Q_2$ or $\log^2 Q_1/Q_2$ require resummation, which sometimes may be interpreted as generalization
 of the Collinear Parton Model (CPM), with the unintegrated (or Transverse Momentum Dependent) PDFs $\Phi_i(x,{\bf q}_T^2,\mu^2)$~\cite{Collins}.
 However, the all-order gauge invariant definition of unPDFs and the hard subprocess, is established only in the kinematic region
  $q_{1,2}^\mp\ll |{\bf q}_{T1,2}|\ll q_{1,2}^\pm=x_{1,2}\sqrt{S}$, where the dependence of $d\hat{\sigma}$ on ${\bf q}_{T1,2}$ can be neglected.

 Another kinematic region, where the unPDFs and hard subprocess can be factorized in a gauge invariant way is the region of Quasi-Multi-Regge
  Kinematics (QMRK), where $q_{1,2}^\mp\ll |{\bf q}_{T1,2}|\sim q_{1,2}^\pm=x_{1,2}\sqrt{S}$ (see~\cite{IFL, QMRKrev} for a review).
  In this region, particles produced in the hard subprocess are highly separated in rapidity from the particles of initial-state radiation (ISR).
  In the QMRK QCD amplitude factorizes into blocks, corresponding to the production of clusters of particles, separated in rapidity.
  Different blocks are connected by the $t$-channel exchange of effective gauge invariant degrees of freedom of QCD -- Reggeized gluons ($R$)
  and Reggeized quarks ($Q$). This factorization is the basis of the $k_T$-factorization formula~\cite{kTf} and the unPDFs satisfy
  Balitsky-Fadin-Kuraev-Lipatov (BFKL) evolution equation.

  Parton Reggeization Approach (PRA) is the hybrid scheme of $k_T$-factorization, which uses Reggeized quarks and gluons to define
  gauge invariant hard-scattering coefficient $d\hat{\sigma}_{PRA}(q_1, q_2)$,
  with $q_{1,2}^\mu=x_{1,2}P_{1,2}^\mu + q_{T1,2}^\mu$, $q_{1,2}^2=-{\bf q}_{1,2}^2=-t_{1,2}$, and Kimber-Martin-Ryskin (KMR) unPDFs~\cite{KMR},
  incorporating the effects of resummation of Sudakov double-logarithms $\log^2 (t/\mu^2)$. Therefore PRA interpolates between TMD factorization
   at small $p_T$ and $k_T$-factorization at high-$p_T$.

  The Leading Order (LO) of PRA works fairly well for the observables, sensitive to the radiation of additional hard partons,
  see e.~g.~\cite{dijets, diphotons}. However, large $K$-factors may appear due to the soft real and virtual emissions, which
  can be captured only by the full next to leading order calculations. Additionally behavior of some observables, such as
  $x_\gamma$-distributions in the associated photoproduction of prompt photon and jet~\cite{gamma-jet}, can not be explained
  in LO of PRA due to kinematic reasons. This motivates the extension of PRA to the Next to Leading Order (NLO) in $\alpha_s$.

  Discussion of real NLO corrections in PRA was started in the Ref.~\cite{diphotons}. At the NLO, real emissions, highly separated
  in rapidity from the central region, may come from both unPDF and NLO hard-scattering matrix element. Corresponding double-counting
  subtraction procedure was introduced in the Ref.~\cite{diphotons}. We postpone for the further studies the consideration of real corrections,
  related with matching of {\it single-scale} observables in PRA and CPM.

  In the present contribution we concentrate on the problem of rapidity divergences, arising in course of calculation of loop corrections
  in the formalism of Lipatov's High energy EFT~\cite{LipatovEFT, LipVyaz}.  The paper is organized as follows. In the Sec.~\ref{sec:EFT}
  the general construction of EFT for Reggeized quarks~\cite{LipVyaz} is outlined and covariant regularization scheme for rapidity
  divergences~\cite{MH-ASV_NLO_qR-vert, MH-ASV_NLO_gR-vert, MH-ASV_2loop-traj} is introduced. In the Sec.~\ref{sec:1-loop} the one-loop corrections
  to the self-energy of Reggeized quark and $\gamma Q q$-scattering vertex are computed. In the Sec.~\ref{sec:comp-QCD} the MRK-asymptotic
  for the NLO correction to the $\gamma\gamma\to q\bar{q}$-amplitude is constructed using the obtained results. After the proper
   {\it localization in rapidity} of the obtained $\gamma Q q$-vertex, the rapidity divergences cancel and the obtained EFT result
    correctly predicts the leading-power term of the full QCD amplitude. Also a few comments concerning the Rapidity Renormalization Group are made.

\section{Effective action and rapidity divergences}\label{sec:EFT}

  Many processes, such as Drell-Yan lepton pair~\cite{DrellYan} or diphoton~\cite{diphotons} hadroproduction are quark-initiated
  and in the framework of PRA it is natural to consider LO and NLO hard-scattering matrix elements with Reggeized quarks in the
  initial state for this processes. The powerful tool to obtain such matrix elements in gauge invariant EFT for Multi-Regge
  processes with quark exchange in $t$-channel, introduced in the Ref.~\cite{LipVyaz}. In this approach, the axis of rapidity
   $y=\log(q^+/q^-)/2$ is sliced into a few regions $y_i\leq y\leq y_{i+1}$, corresponding to the clusters of produced particles,
   and a separate copy of the full QCD-Lagrangian is defined in each of them:
  \[
\hspace*{-5mm}  L_{QCD}(A_\mu, \psi_q)=-\frac{1}{2}\tr\left[G_{\mu\nu}G^{\mu\nu}\right]+\sum\limits_{q=1}^{n_F}\bar{\psi}_q (i\hat{D}-m_q)\psi_q,
  \]
  where $\hat{D}=\gamma_\mu D^\mu$, $D_\mu=\partial_\mu+ig_s A_\mu$ -- covariant derivative, containing the gluon field $A_\mu=A_\mu^a T^a$, where $T^a$ are (Hermitian) generators of fundamental representation of $SU(N_c)$ gauge group, $g_s$ is the Yang-Mills coupling constant and $G_{\mu\nu}=-i\left[D_\mu,D_\nu\right]/g_s$ is the non-Abelian field-strength tensor. The full effective Lagrangian for the QMRK processes with quark exchanges in $t$-channel reads:
  \begin{eqnarray}
\hspace*{-5mm} & L_{eff}= L_{kin}(Q_+, Q_-) + \sum\limits_i \left[ L_{QCD}(A_\mu^{[y_i, y_{i+1}]}, \psi_q^{[y_i, y_{i+1}]})\right. \nonumber \\
\hspace*{-5mm} & \left. + L_{ind}(A_\mu^{[y_i, y_{i+1}]}, \psi_q^{[y_i, y_{i+1}]}, Q_+, Q_-)  \right],   \label{eq:Leff}
  \end{eqnarray}
where the label $[y_i, y_{i+1}]$ denotes that the real part of rapidity $y$ of the momentum modes of the corresponding field is
constrained: $y_i\leq y \leq y_{i+1}$. Quarks and gluons, produced in the different intervals in rapidity, communicate via the
exchange of Reggeized quarks with the kinetic term:
\begin{equation}
L_{kin}(Q_+, Q_-)=2\left( \overline{Q}_+ i\hat{\partial} Q_- + \overline{Q}_- i\hat{\partial} Q_+ \right), \label{eq:kinQ}
\end{equation}
and the fields $Q_+$ or $Q_-$ are subject to the following kinematic constraint:
\begin{eqnarray}
  \partial_\pm Q_\mp =0,\ \partial_\pm \overline{Q}_{\mp} =0, \label{eq:kin_cons_Q1} \\
  \hat{n}^\pm Q_{\mp} =0,\ \overline{Q}_{\pm} \hat{n}^{\mp}=0. \label{eq:kin_cons_Q2}
\end{eqnarray}
  where the light-like vectors $n_\pm^\mu$ are pointing along the momenta of highly energetic initial-state particles,
  $n_+n_-=2$ and $\partial_\pm=n_\pm^\mu \partial_\mu=2 \partial/\partial x_\mp$, $x_\pm=x^\pm=n_\pm x=x^0\pm x^3$
  in the center of mass frame of initial state.  Conditions (\ref{eq:kin_cons_Q1}, \ref{eq:kin_cons_Q2}) are equivalent
  to the requirement of QMRK for the particles in the final state. Consequently, the propagator of Reggeized quark
   $i \hat{P}_\pm \hat{q} \hat{P}_\pm/q^2$ contains the projectors $\hat{P}_\pm=\hat{n}_\mp\hat{n}_\pm /4 $ ensuring
    the constraint (\ref{eq:kin_cons_Q2}). The fields $Q_\pm$ are gauge invariant, and the corresponding {\it induced} interaction term reads:
  \begin{equation}
 \hspace*{-5mm} L_{ind}= - \overline{Q}_- i\hat{\partial} \left( W^\dagger [A_+] \psi \right) - \overline{Q}_+ i\hat{\partial} \left( W^\dagger [A_-] \psi \right) + \Hc, \ \label{eq:Lind}
  \end{equation}
   where $W[A_\pm]=P\exp\left[(-ig_s/2) \int\limits_{-\infty}^{x_{\mp}} dx'_{\mp} A_\pm (x_{\pm},x'_{\mp},{\bf x}_T)  \right]$
   is the Wilson line stretched along the light-cone, which can be expanded perturbatively as:
  \begin{equation}
\hspace*{-5mm}  W[A_\pm]= 1-ig_s(\partial_\pm^{-1} A_\pm)+ (-ig_s)^2 (\partial_\pm^{-1} A_\pm\partial_\pm^{-1} A_\pm)+... \
  \end{equation}
  leading to the infinite number of nonlocal interaction vertices of Reggeized quark ($Q$), Yang-mills quark ($\psi$) and $n$
  Yang-Mills gluons ($A_\mu$). In particular, the Fadin-Sherman~\cite{FadinSherman, FadinBogdan} $gQq$-vertex and $ggQq$-vertex reads:
  \begin{eqnarray}
 \hspace*{-10mm} &  ig_sT^a\gamma^{(\pm)}_\mu(q,k)=ig_s T^a \left(\gamma_\mu + \hat{q} \frac{n^\mp_\mu}{k^\mp} \right), \label{eq:gQq-vert} \\
\hspace*{-10mm} &  \gamma^{(\pm)}_{\mu\nu}(q,k_1,k_2)=ig_s^2 (n^\mp_{\mu_1} n^\mp_{\mu_2}) \hat{q} \left[ \frac{T^{a_1} T^{a_2}}{k_1^\mp(k_1+k_2)^\mp} + (1\leftrightarrow 2) \right] \label{eq:ggQq-vert}
  \end{eqnarray}
  where $q$ and $k_i$ are the (incoming) momenta of Reggeized quark and of gluons, respectively. The induced vertices contain nonlocal
  factors $1/k^\pm$ which  in certain kinematics can lead to integrals, which are ill-defined in the dimensional regularization.
  To demonstrate this, let's study the scalar integral which will appear in the computation of the self-energy of the Regeized quark:
  \begin{equation}
\hspace{-5mm}  B_0^{[+-]}({\bf p}_T)=\int\frac{d^D q}{i\pi^{D/2}} \frac{1}{q^2 (p-q)^2 [q^+] [q^-] },
  \end{equation}
  where, following the approach of Ref.~\cite{MH_Pole-prescription}, we have assigned the Principal Value (PV) $i\varepsilon$-prescription to $1/q^\pm$ poles:
\begin{equation}
\frac{1}{[q^\pm]}=\frac{1}{2}\left(\frac{1}{q^\pm+i\varepsilon} + \frac{1}{q^\pm-i\varepsilon}\right),\label{eq:VP}
\end{equation}
 while the standard $i\varepsilon$-prescription for Feynman propagators is kept implicit.

 Due to constraint (\ref{eq:kin_cons_Q1}) the external momentum $p$ is purely transverse $p^+=p^-=0$.  Transferring to the integration
 over light-cone components $d^D q= dq^+ dq^- d^{D-2} {\bf q}_T/2$ and substituting the anzats $q^\pm=\xi^\pm e^{\pm y}\sqrt{|q^2+{\bf q}_T^2|}$, $\xi^+\xi^-=\sgn(q^2+{\bf q}_T^2)$, one finds, that integral in rapidity is factorized:
  \begin{eqnarray}
\hspace*{-7mm} & B_0^{[+-]} = \int\limits_{y_i}^{y_{i+1}} dy \int \frac{d^{d-2}{\bf q}_T}{2\pi^{D/2} i} \int\limits_{-\infty}^{+\infty} \frac{d q^2}{(q^2+{\bf q}_T^2-i\varepsilon) (q^2+i\varepsilon)}\times \nonumber \\
\hspace*{-10mm} &  \frac{1}{(q^2+2{\bf p}_T {\bf q}_T - {\bf p}_T^2 + i\varepsilon)} = \int \frac{d^{D-2}{\bf q}_T}{\pi^{(D-2)/2}} \frac{(y_{i+1}-y_i)}{{\bf q}_T^2 ({\bf p}_T-{\bf q}_T)^2}. \label{eq:B0pm_y}
  \end{eqnarray}

 The typical two-dimensional IR-divergent integral, contributing to the LO Regge trajectories of gluon~\cite{LipatovEFT} and
 quark~\cite{FadinSherman, FadinBogdan}, appears in the last expression in front of $(y_{i+1}-y_i)$-divergence.

 The regularization of EFT~\cite{LipatovEFT,LipVyaz} by explicit cutoffs in rapidity is very physically clear, the $y_i$-dependence
 should cancel between the contributions of neighboring regions in rapidity order-by-order in $\alpha_s$, building up a corresponding $\log^n s$-factor. However, in practical applications to the multiscale Feynman integrals, the cutoff regularization is quite inconvenient. We would like to keep manifest Lorentz-covariance of the formalism and to be able to make shifts of integration momenta, also we would like to preserve the gauge invariance of EFT~\cite{LipatovEFT,LipVyaz} after regularization. Different versions of the analytic regularization for rapidity divergences can be found in the literature on Soft-Collinear Effective Theory (see e. g.~\cite{SCET_RRG}), however the proof of gauge invariance of regularized expressions is quite involved in this case. In the present study we use, {\it covariant regularization} developed in Refs.~\cite{MH-ASV_NLO_qR-vert, MH-ASV_NLO_gR-vert, MH-ASV_2loop-traj}. Following this approach, we shift the $n_\pm$-vectors in the induced interactions (\ref{eq:Lind} -- \ref{eq:ggQq-vert}) slightly off the light-cone:
 \begin{eqnarray}
\hspace*{-8mm} & n_\pm^\mu \to \tilde{n}_\pm^\mu= n_\pm^\mu + e^{-\rho} n_\mp^\mu ,\  k_\pm \to \tilde{k}_\pm= k_\pm + e^{-\rho} k_\mp ,  \label{eq:Npm} \\
\hspace*{-8mm} & \tilde{n}^2_\pm=4 e^{-\rho},\ \tilde{n}_+\tilde{n}_-=2(1+e^{-\rho}),\label{eq:Npm2}
 \end{eqnarray}
 where $\rho\gg 1$ is the regulator variable. In the next section we will list all single-scale 1,2 and 3-point one-loop scalar integrals,
 necessary for the computations, and comment on the appearing rapidity divergences.

\section{NLO corrections to the $\gamma Q q$-vertex and Reggeized quark self-energy}\label{sec:1-loop}

  First, we list the one-loop Feynman integrals appearing in the calculation. We will work in $D=4-2\epsilon$-dimensions to regularize the
   UV and IR divergences and add the factor $r_\Gamma=\Gamma^2(1-\epsilon)\Gamma(1+\epsilon)/\Gamma(1-2\epsilon)$ to the denominators of
   all integrals to be compatible with the notation of Ref.~\cite{K_Ellis} for ordinary scalar integrals.

  The following ``tadpole'' integral is nonzero:
  \[
\hspace*{-7mm}  A_0^{[+]}(p)=\int \frac{[d^Dq]}{(p-q)^2[\tilde{q}^+]}=  \frac{\tilde{p}^+}{\cos (\pi\epsilon)} \frac{ e^{\rho(1-\epsilon)}}{2\epsilon(1-2\epsilon)}\left\lbrace \frac{\mu}{\tilde{p}^+} \right\rbrace^{2\epsilon}, 
  \]
  where $[d^Dq]= (\mu^2)^\epsilon d^Dq/(i\pi^{D/2} r_\Gamma)$ and $\left\lbrace \frac{\mu}{k} \right\rbrace^{2\epsilon}=\frac{1}{2}\left[ \left( \frac{\mu}{k-i\varepsilon} \right)^{2\epsilon} + \left( \frac{\mu}{-k-i\varepsilon} \right)^{2\epsilon} \right]$. Integral $A_0^{[+]}$ contains the {\it power} $e^\rho$ and {\it weak-power} $e^{-\epsilon\rho}$ divergences. Here and in the following, the $I^{[-]}$ integral can be obtained by the $(+\leftrightarrow -)$ substitution int the $I^{[+]}$ integral.

  The asymptotic expansion for $\rho\gg 1$ of the following ``bubble'' integral was computed using Mellin-Barnes representation~\cite{Smirnov}:
  \[
\hspace*{-7mm}   B_0^{[+]}(p)=\int \frac{[d^Dq]}{q^2 (p-q)^2[\tilde{q}^+]} =  \frac{1}{\tilde{p}^+} \frac{e^{-\epsilon\rho}}{ \cos (\pi \epsilon)} \frac{1}{2\epsilon^2} \left\lbrace \frac{\mu}{\tilde{p}^+} \right\rbrace^{2\epsilon},
  \]
  and all terms $\sim e^{-n\rho/2}$ for $n\geq 1$ are neglected. In course of calculation, the leading $e^{\rho/2}$ power divergence,
  which can be found in the Ref.~\cite{MH-ASV_NLO_gR-vert}, has canceled due to the PV-prescription (\ref{eq:VP}). For the case $p^+=p^-=0$
  integral $B_0^{[+]}({\bf p}_T)=0$ because Mellin-Barnes representation starts with terms $\sim e^{-\rho/2}$.

  In course of tensor reduction, one has to do the momentum-shifts $(q-p)^2(q-k)^2\to q^2 (k+p-q)^2$, where $\tilde{p}^+\sim e^{-\rho}$,
  then the following integral appears:
  \[
\hspace{-8mm}  \int\frac{[d^Dq] }{q^2(p+k-q)^2 [\tilde{q}^+-\tilde{p}^+]}=B_0^{[+]}(p+k)-\frac{\tilde{p}^+}{\tilde{k}^+} B_0^{[+]}(p),
  \]
  up to power corrections in $e^{-\rho}$.

  The integral with two eikonal propagators:
  \begin{equation}
  B_0^{[+-]}({\bf p}_T)=\frac{1}{{\bf p}_T^2}\left(\frac{\mu^2}{{\bf p}_T^2}\right)^\epsilon \frac{i\pi - 2\rho}{\epsilon},
  \end{equation}
  contain the {\it logarithmic} rapidity divergence $\sim \rho$ and can be extracted from the results of Ref.~\cite{MH-ASV_NLO_qR-vert}.
  It is easy to see, that logarithmic divergence $\sim\rho$ is equivalent to $(y_{i+1}-y_i)$ in Eq. (\ref{eq:B0pm_y}), and also the imaginary
   part appears, which is related to the imaginary part of the forward scattering amplitude in full QCD.

  The ``triangle'' integral with one eikonal propagator:
  \[
  C_0^{[+]}(k^+,{\bf p}_T^2)=\int \frac{[d^Dq]}{q^2 (p-q)^2 (p+k-q)^2 [\tilde{q}^+]},
  \]
  where $k^2=k^-=0$, $p^2=-{\bf p}_T^2$, $(p+k)^2=0$ is obtained using Mellin-Barnes representation. The result coincides
   with the result of Ref.~\cite{MH-ASV_NLO_gR-vert}:
   \begin{eqnarray*}
 \hspace{-5mm} C_0^{[+]}(k^+, {\bf p}_T^2) = \frac{1}{k^+ {\bf p}_T^2}\left( \frac{\mu^2}{{\bf p}_T^2} \right)^\epsilon \frac{1}{\epsilon} \left[  \rho -i\pi + \log \frac{(k^+)^2}{{\bf p}_T^2} +  \right. \nonumber \\
 \left. \psi(1+\epsilon) + \psi(1) - 2\psi(-\epsilon) \right],
   \end{eqnarray*}
   also contains the logarithmic rapidity divergence and imaginary part.  These are all one-loop rapidity-divergent
    integrals with one virtuality scale, which appear in EFT~\cite{LipatovEFT,LipVyaz}.

   The tensor reduction of one-loop integrals follows the usual Passarino-Veltman procedure, except that vectors
    $\tilde{n}_\pm$ should be added to the anzats for the tensor structure. For example, solution for the integral $B^{[+]}(p)$
    with one factor of $q_\mu$ in the numerator contains vectors $p^\mu$ and $\tilde{n}^\mu_+$, and the corresponding Gram
     determinant is equal to $\Delta=(\tilde{p}^+)^2-4e^{-\rho} p^2$, so that in the case when $\tilde{p}^+\sim e^{-\rho}$,
     the coefficient in front of $\tilde{n}_+^\mu$ will be proportional to $e^\rho$. Because of this power divergences one has to keep
     the exact $\rho$-dependence of coefficients in front of scalar integrals, arising from Eq.~(\ref{eq:Npm2}), up to the moment when
     all power divergences cancel out. Then, the terms suppressed by powers of $e^{-\rho}$ can be dropped.

   \begin{figure}[ht]
   \begin{center}
   \begin{tabular}{ccc}
   \includegraphics[width=0.35\textwidth]{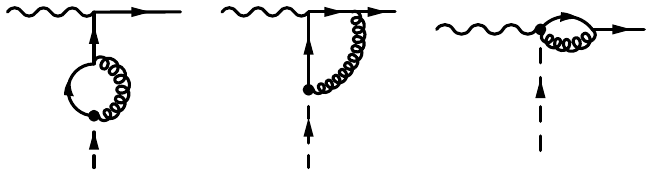} &
   \includegraphics[width=0.07\textwidth]{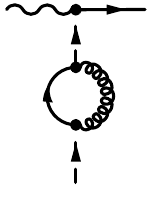} \\
   (a) & (b)
   \end{tabular}
   \caption{Diagrams, contributing to $\gamma Q q$-vertex at one-loop -- (a) and subtraction term for the one-loop $\gamma Qq$-vertex -- (b).
   Reggeized quark is denoted by dashed line.   \label{fig1}}
   \end{center}
   \end{figure}

   Now we are in a position to present the results of computation of self-energy of the Reggeized quark and the $\gamma Qq$-vertex at one-loop.
   The one-loop Reggeized quark self-energy reads:
   \begin{eqnarray*}
   \hspace*{-7mm}  & \hat{\Sigma}_1({\bf p}_T)=\frac{C_F\bar{\alpha}_s}{4\pi} (-i)\int [d^Dq] \frac{\gamma_\mu^{(-)}(-p,q) (\hat{p}-\hat{q}) \gamma_\mu^{(+)}(p,-q)}{q^2(p-q)^2},
   \end{eqnarray*}
 where $\bar{\alpha}_s=\alpha_s (4\pi/\mu^2)^\epsilon r_\Gamma$ and $C_F=(N_c^2-1)/(2N_c)$. And the result is:
   \begin{eqnarray}
 \hspace{-7mm} & \hat{\Sigma}_1({\bf p}_T)=(-i\hat{p}) \frac{C_F \bar{\alpha}_s}{4\pi} \left(\frac{\mu^2}{{\bf p}_T^2}\right)^\epsilon \left[ \frac{{2\rho}-i\pi}{\epsilon} + \left(\frac{1+\epsilon}{1-2\epsilon}\right) \frac{1}{\epsilon} \right].
   \end{eqnarray}
   Apart from the $\rho$-divergence, leading to the quark Reggeization, the self-energy graph also contains the UV-pole.
   The integrals containing power rapidity divergences do not appear.

   The Feynman diagrams contributing to the one-loop correction to $\gamma Qq$-vertex are shown in the Fig.~\ref{fig1}(a).
   We will consider the kinematics, when high-energy photon with large $k^+$ momentum component scatters on the Reggeized
   quark with virtuality $p^2=-t_1$, $t_1>0$ and produces massless quark. To obtain the result free from weak-power rapidity
   divergences $\sim e^{\pm \epsilon \rho}$ one have to take into account the corresponding high-energy projector $\hat{P}_\pm$
   in the propagator of Reggeized quark.  Then scalar integrals, described above, appear in the result of tensor reduction,
   but the coefficients in front of integrals $A_0^{[+]}$ and $B_0^{[+]}$ are of order $O(e^{-2\rho})$ or $O(e^{-\rho})$, and,
   therefore, the result is free from power and weak-power rapidity divergences. Only logarithmic rapidity divergence, originating
   from integral $C_0^{[+]}$ is left. After neglecting all terms suppressed by powers of $e^{-\rho}$ one can expand the result
   in $\epsilon$, the expanded result reads:
   \begin{eqnarray}
\hspace{-7mm}   \hat{\Gamma}_1^\mu=\frac{2}{t_1}\hat{\Delta}_0^\mu +\hat{\Gamma}_0^\mu\left[-\frac{1}{\epsilon^2}-\frac{L_1}{\epsilon} +(\rho-i\pi) \left(\frac{1}{\epsilon}+L_1\right) +\right. \nonumber \\
\hspace{-7mm} \left. \frac{2L_2}{\epsilon} - \left(\frac{1}{\epsilon}+L_1+3\right) +2L_1L_2 - \frac{L_1^2}{2} + \frac{\pi^2}{2}  \right]  + O(\epsilon),\label{eq:unsVert}
   \end{eqnarray}
   where the overall factor $\bar{\alpha}_sC_F/(4\pi)$ is omitted, $L_1=\log\left(\frac{\mu^2}{t_1}\right)$, $L_2=\log\left(\frac{k^+}{\sqrt{t_1}}\right)$ and the Dirac structures $\hat{\Gamma}_0$ and $\hat{\Delta}_0$ have the form:
   \begin{eqnarray*}
   \left(\hat{\Gamma}_0 \right)_\mu=i ee_q \bar{u}(p+k)\gamma^{(-)}_\mu(p,k)\hat{P}_-,\\
   \hat{\Delta}_0^\mu=i ee_q \left(p^\mu - \frac{t_1}{2k^+} n_+^\mu\right) \left( \bar{u}(p+k) \hat{k} \hat{P}_- \right),
   \end{eqnarray*}
   where $e$ is the QED coupling constant and $e_q$ is the quark electric charge. It is important to notice, that the obtained result
   obeys the QED Ward identity $k_\mu \hat{\Gamma}_1^\mu=0$ independently on $t_1$. In fact, the full $\rho$-dependent result, before
   omitting the power-suppressed terms, also obeys Ward identity, and the third diagram in the Fig.~\ref{fig1}(a) is particularly
   important for this. In Ref.~\cite{MH-ASV_NLO_gR-vert} contributions of diagrams of this topology where nullified by the gauge-choice
   for external gluons.

  Covariant rapidity regulator breaks-down explicit separation of the contributions of different regions in rapidity.
  Regularization of the $1/{q^+}$-pole in the vertex correction corresponds to the cutoff of the loop momentum at some
  large negative rapidity $y_1(\rho)<0$, while regularization of $1/q^+$ and $1/q^-$-poles in the self-energy correction
  constrains the rapidity of gluon in the loop from both sides: $y_1(\rho)<y<y_2(\rho)$. Clearly, if we just add the Reggeon
  self-energy graph to the vertex correction, the contribution of region $y_1(\rho)<y<y_2(\rho)$ will be double-counted and there
   is no reason to expect the cancellation of $\sim\rho$ divergences. Before adding all NLO corrections one have to ``localize''
   the vertex correction in rapidity by subtraction of the corresponding central-rapidity
   contribution~\cite{MH-ASV_NLO_qR-vert, MH-ASV_NLO_gR-vert, MH-ASV_2loop-traj}. The diagrammatic representation for the corresponding
    subtraction term is shown in the Fig.~\ref{fig1}(b), and after $\epsilon$-expansion it has the form:
  \begin{equation}
 \hspace{-7mm} \delta\hat{\Gamma}_1^\mu=\hat{\Gamma}_0^\mu\left[ (2\rho-i\pi) \left(\frac{1}{\epsilon}+L_1 \right) + \left( \frac{1}{\epsilon}+L_1+3 \right)\right] + O(\epsilon). \label{eq:sub_term}
  \end{equation}
  Apart from the rapidity divergence, the subtraction term contains also $1/\epsilon$ UV-pole, associated with the UV-divergence
   of the ordinary quark propagator.

\section{Comparison with QCD. Rapidity Renormalization Group.}\label{sec:comp-QCD}
  The results obtained above can be used to construct the Regge asymptotics of $\gamma\gamma\to q\bar{q}$ scattering amplitude at one-loop.
  To facilitate the comparison with the full QCD result we will consider the scalar quantities -- the real and imaginary parts of the interference
  term between LO tree-level and one-loop amplitude. In the Regge limit, the parameter $\tau=-t/s$ is small, and the interference term can be
  expanded as follows:
  \begin{equation}
 \hspace{-5mm}  \sum\limits_{\rm col.,\ spin} M_{1-loop} M^*_{LO}  = \frac{1}{\tau}C_{HE}^{(-1)} + C_{HE}^{(0)} + O(\tau),
  \end{equation}
  where the overall factor $(8\cdot (e e_q)^4 N_c)\frac{C_F \bar{\alpha}_s}{4\pi}$ is omitted and the EFT~\cite{LipVyaz} predicts the coefficient
   in front of the leading power -- $C_{HE}^{(-1)}$.

   After the subtraction of the contribution (\ref{eq:sub_term}) from the corrections to the $\gamma Q_+ q$ and $\gamma Q_- q$ scattering vertices,
   all rapidity divergences cancel at the order $\alpha_s$, and the results for the real and imaginary parts of the coefficient $C_{HE}^{(-1)}$
   looks as follows:
   \begin{eqnarray}
\hspace*{-10mm} &  \re C_{HE}^{(-1)}=-\frac{2}{\epsilon^2} - \frac{2}{\epsilon} \log \frac{\mu^2}{(-t)}+ \left( 1+2\log\frac{1}{\tau} \right)\left(\frac{1}{\epsilon}+\log \frac{\mu^2}{(-t)}\right)\nonumber\\
\hspace*{-10mm} &- \left[ 3-\pi^2 +\log ^2 \frac{\mu^2}{(-t)}+4\log\frac{1}{\tau}\right] , \label{eq:CHE_re}\\
\hspace*{-10mm} &  \im C_{HE}^{(-1)}=-\pi \left( \frac{1}{\epsilon} + \log \frac{\mu^2}{(-t)}  -2  \right) . \label{eq:CHE_im}
   \end{eqnarray}

\begin{figure}
\begin{center}
\begin{tabular}{cc}
\includegraphics[width=0.17\textwidth]{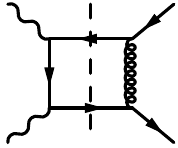} & \includegraphics[width=0.17\textwidth]{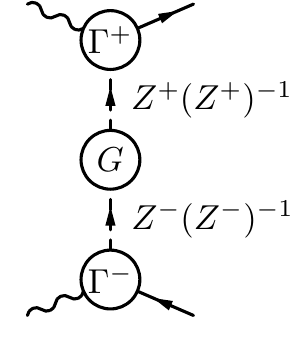} \\
(a) & (b)
\end{tabular}
\end{center}
\caption{The cut contributing to the imaginary part of $\gamma\gamma\to q{\bar q}$ amplitude -- (a).
Contribution with the exchange of one Reggeon in $t$-cahnnel -- (b).\label{fig2}}
\end{figure}

   We have performed the computation of real and imaginary parts of $C_{HE}^{(-1)}$ in full QCD in dimensional regularization,
   using the \texttt{FeynArts}~\cite{FeynArts} and \texttt{FeynCalc}~\cite{FeynCalc} packages. The obtained results agree with
   the expressions (\ref{eq:CHE_re}), (\ref{eq:CHE_im}), derived within the EFT~\cite{LipVyaz}, providing the nontrivial test
   of the self-consistency of the formalism. In particular, the imaginary part (\ref{eq:CHE_im}) originates from the box diagram
   with quark and gluon exchange in the $t$-channel (Fig.~\ref{fig2}(a)).

   The EFT~\cite{LipatovEFT,LipVyaz} provides the efficient tool for resummation of high-energy logarithms $\sim\log 1/\tau$. Assuming,
   that after appropriate localization of all loop corrections in rapidity, logarithmic rapidity divergences cancel order by order in $\alpha_s$
   within the subset of diagrams with exchange of one Reggeon in $t$-channel (Fig.~\ref{fig2}(b)), one can introduce the multiplicative
   renormalization to remove them from corrections to the $\gamma Q_\pm q$-vertices and propagator of the Reggeized quark~\cite{MH-ASV_2loop-traj}:
  \begin{eqnarray*}
& \Gamma^\pm_R\left(\frac{k^\pm}{M^\pm}, \frac{t}{\mu^2} \right)= Z^\pm\left(\rho, \frac{t}{\mu^2} ,\frac{M^\pm}{\sqrt{\mu^2}}\right) \Gamma^\pm\left( \rho, \frac{k^\pm}{\sqrt{\mu^2}}, \frac{t}{\mu^2} \right),\\
& G_R\left(\frac{M^+}{\sqrt{\mu^2}}, \frac{M^-}{\sqrt{\mu^2}}, \frac{t}{\mu^2} \right) = (Z^+Z^-)^{-1} G\left(\rho, \frac{t}{\mu^2} \right),
\end{eqnarray*}
  where the scales $M^\pm$ are introduced to parametrize the finite ambiguity in the definition of subtraction scheme for $\rho$-divergences,
  related with the possible redefinition of the regulator variable, which doesn't spoil the $t$-channel factorization of the amplitude: $\rho \to \rho - 2\log M^\pm/\sqrt{\mu^2}$. The bare quantities do not depend on $M^\pm$, and therefore the renormalized quantities obey the Rapidity Renormalization Group (RRG) equations:
    \[
  \frac{\partial\log G_R}{\partial\log M^\pm}=\omega(t),\  \frac{\partial\log \Gamma^\pm_R}{\partial\log M^\pm}=-\omega(t),
  \]
  where the corresponding anomalous dimension  $\omega(t)=\lim\limits_{\rho\to\infty} \frac{(-1)}{Z^\pm}\frac{\partial Z^\pm}{\partial \log M^\pm}$
  is nothing but the quark Regge trajectory. At one loop we have:
  \begin{eqnarray*}
\hspace*{-10mm} & Z^\pm = 1+ \frac{\bar{\alpha}_s}{4\pi}C_F \left(\rho - 2\log \frac{M^\pm}{\mu}\right)\left(\frac{1}{\epsilon}+\log\frac{\mu^2}{(-t)}\right)+O(\bar{\alpha}_s^2), \\
\hspace*{-10mm} &\omega(t)=\frac{\bar{\alpha}_s}{2\pi} C_F \left( \frac{1}{\epsilon}+ \log\frac{\mu^2}{(-t)} \right) + O(\bar{\alpha}_s^2,\epsilon),
  \end{eqnarray*}
  in full consistency with the results present in the literature~\cite{FadinSherman, FadinBogdan}.

  To remove the large logarithms $\log k^\pm/M^\pm$ from vertex corrections, one can set $M^+=k_1^+$ and $M^-=k_2^-$,
  then it is possible to resum the corrections enhanced by $\log (M^+M^-)= \log s$ by solving the RRG equation for the Reggeon propagator:
  \[
  G_R(t,s)=\left(\frac{s}{M_0^+M_0^-}\right)^{\omega(t)} G_R(t,M_0^+M_0^-),
  \]
  where one can take the Born propagator of the Reggeized quark as the initial condition for the evolution $G_R(t,M_0^+M_0^-)$
  at the starting scale $M_0^+M_0^-\sim -t$. In this way, the Reggeization of quark, as well as of a gluon, can be understood within
  the context of EFT~\cite{LipatovEFT,LipVyaz} with covariant rapidity regulator.

\section*{Conclusions}

  The results obtained above show, that the gauge invariant effective action for high energy processes in QCD~\cite{LipatovEFT, LipVyaz}
  together with the covariant rapidity regulator, proposed in
  the Refs.~\cite{MH-ASV_NLO_qR-vert, MH-ASV_NLO_gR-vert, MH-ASV_2loop-traj, MH_Pole-prescription} is a powerful tool to resum radiative
  corrections enhanced by the high energy logarithms $\log s$ or $\log 1/x$ in Leading Logarithmic Approximation and beyond.
  This tool can be used in the fenomenological calculations in PRA at NLO, but the technique to calculate the integrals with
  more than one virtuality scale should be developed.

\section*{Acknowledgments}

  Authors are grateful to B. A. Kniehl, G.~Chachamis and A. Sabio Vera for clarifying communications concerning
  the computation of scalar integrals. The work was supported in part by RFBR grant No. 14--02--00021 and the grant
  of Ministry of Education and Science of Russian Federation No. 1394.

\nocite{*}
\bibliographystyle{elsarticle-num}
\bibliography{martin}



\end{document}